\documentclass[preprint,showpacs,showkeys,preprintnumbers,amsmath,amssymb,superscriptaddress]{revtex4}
\usepackage{graphicx}
\usepackage{dcolumn}
\usepackage{bm}

\begin{document}
\title{Entropic force approach in a noncommutative charged black hole and the equivalence principle}

\author{S. Hamid Mehdipour}

\email{mehdipour@liau.ac.ir}

\affiliation{Islamic Azad University, Lahijan Branch, P. O. Box
1616, Lahijan, Iran}

\author{A. Keshavarz}

\email{arash_batty2004@yahoo.com}

\affiliation{Islamic Azad University, Central Tehran Branch, P. O.
Box 13185-768, Tehran, Iran}

\date{\today}

\begin{abstract}
Recently, Verlinde has suggested a novel model of duality between
thermodynamics and gravity which leads to an emergent phenomenon for
the origin of gravity and general relativity. In this paper, we
investigate some features of this model in the presence of
noncommutative charged black hole by performing the method of
coordinate coherent states representing smeared structures. We
derive several quantities, e.g. temperature, energy and entropic
force. Our approach clearly exhibits that the entropic force on a
smallest fundamental cell of holographic surface with radius $r_0$
is halted. Accordingly, we can conclude that the black hole remnants
are absolutely inert without gravitational interactions. So, the
equivalence principle of general relativity is contravened due to
the fact that it is now possible to find a difference between the
gravitational and inertial mass. In other words, the gravitational
mass in the remnant size does not emit any gravitational field,
therefore it is experienced to be zero, contrary to the inertial
mass. This phenomenon illustrates a good example for a feasible
experimental confirmation to the entropic picture of Newton's Second
law in very short distances.
\end{abstract}

\pacs{04.70.Dy, 04.50.Kd, 05.70.-a, 02.40.Gh} \keywords{Black Hole
Thermodynamics, Noncommutative Geometry, Holographic Screens,
Entropic Force}

\maketitle

The thermal emission from black holes has primarily proved the firm
relations between thermodynamics and the gravitational treatment of
black holes \cite{haw}. It seems that one noticeable evidence for
the essence of gravity is perceived from the precise inspection of
black holes thermodynamics because it is feasible to prepare a
physical similarity between spacetimes comprising horizons and the
concepts of temperature or entropy. Moreover, since a quantum theory
of gravity tells us that the black hole entropy could be connected
to a number of microscopic states, the meticulous assessment of
black hole entropy, due to possessing the significant concepts for a
yet to be formulated theory of quantum gravity, can be important to
construct a full theory of quantum gravity. This is the principal
reason to investigate the origin of the black hole entropy at a
fundamental level. A few years ago, Jacobson demonstrated that the
Einstein equations are attained from the thermodynamics laws
\cite{jac}. In 2009, Padmanabhan performed the debate of
equipartition energy of horizons to provide a thermodynamic outlook
on gravity \cite{pad1}. In 2010, Verlinde proposed a new viewpoint
of duality between thermodynamics and gravity which has emerged,
{\it as an entropic force}, by information changes linked to the
locations of physical objects \cite{ver}. This hypothesis indicates
that gravitational interaction emerges from the statistical
treatment of microscopic degrees of freedom encrypted on a
holographic surface and can be illustrated as a type of entropic
force associated with the information which is accumulated on the
holographic screens. The concept of entropic force in various
situations has been scrutinized by many authors \cite{ent}. In
addition, there are some comments on Verlinde's entropic gravity
approach which point out short comings of this approach as well as
open challenges \cite{comm}, they provide some unanswered questions
about the issue.

Considering the gravitational force is entropic-like and entropy
combines the emergent view of gravity with the fundamental
microstructure of a quantum spacetime, it is necessary to take into
consideration the microscopic scale influences by applying precise
implements like noncommutative gravity to illustrate the microscopic
structure of a quantum spacetime, in Verlinde's proposal. In 2005,
Nicolini {\it et al} \cite{nic1} in a novel method to noncommutative
gravity, which is established upon coordinate coherent state
approach \cite{sma}, recuperated the short distance treatment of
point-like structures. This method is the so-called
{\it{noncommutative geometry inspired model}}. They have clarified
that the evaporation procedure of black hole must be ceased when the
black hole gets to a minimal nonzero mass called a stable
Planck-sized remnant. This residual mass of the black hole arises
from the existence of minimal observable length. The smallness of
this scale would infer that noncommutativity effects can be
apprehended just in extreme energy phenomena. The majority of the
phenomenological analysis of the noncommutativity scenarios have
assumed that the noncommutative energy scale cannot lie far above
the TeV scale \cite{hin}. On the other hand, the fundamental Planck
scale in models with extradimensions \cite{ant} can be adjacent to
current particle physics experiments \cite{ade}; it is feasible to
set the noncommutativity effects in a TeV regime. In addition, some
type of divergencies which are revealed in general relativity can be
removed in the noncommutativity framework. To illustrate more
features, see \cite{nic2} and the references included.

In this paper, we use noncommutative geometry inspired model to
unite the microscopic structure of spacetime with the entropic
interpretation of gravity due to the fact that the idea of entropy
has a substantial connection to the quantum spacetime structures. We
investigate the entropic force method in the presence of
noncommutative Reissner-Nordstr\"{o}m (RN) black hole to observe the
possible novel phenomena due to the effects of smearing of the
particle mass and charge. Our studies declare that there is a
contravention of the equivalence principle of general relativity
when we include the noncommutativity corrections in our
calculations. The equivalence principle (EP) is a perceivable
foundation for general relativity claiming that one cannot find a
difference between a uniform acceleration and a gravitational field
in a locally frame of reference. We will show that an evident
violation of the EP exists. This enables one to locally mark a
difference between the gravitational and inertial mass. Thus, an
inherent trait for the fundamental microstructure of a quantum
spacetime like noncommutative gravity can lead to a violation of the
EP. In a recent paper \cite{meh2}, we addressed several issues of
entropic nature of gravity as proposed by Verlinde in the framework
of noncommutative geometry inspired model for the Schwarzschild
black hole. As a result of spacetime noncommutativity, Einstein
equations in vacuum have a Schwarzschild black hole solution which
has a mass distributed in a region in place of a mass completely
localized at a point. We implemented two different distributions:
(a) Gaussian and (b) Lorentzian, in order to derive several
quantities, e.g. temperature, energy and entropic force. Both mass
distributions prepared the similar quantitative aspects for the
entropic force. In this setup, if one considered the screen radius
less than the radius of smallest holographic surface, one would
encounter some unusual dynamical features leading to negative
entropic force, i.e. gravitational repulsive force and it is worth
of mentioning: at this regime either our analysis is not the proper
one, or non-extensive statistics should be employed. Therefore, we
will henceforth apply the circumstance that the screen radius is
bigger than the radius of the smallest holographic surface.

In the end, it should be noted that there have been other proposals
which contravene the EP such as the quantum phenomenon of neutrino
oscillations \cite{gas}, comparing Hawking radiation to Unruh
radiation \cite{dug}, and examination of entropic picture of
Newton's second law for the case of circular motion
\cite{dun} (somewhat more related to the present work).\\

The method we consider here is to search for a static,
asymptotically flat, spherically symmetric, minimal width, Gaussian
distribution of mass and charge whose noncommutative size is made by
the parameter $\sqrt{\theta}$. For this purpose, we are going to
exhibit the mass and charge distributions by a smeared delta
function $\rho$ (see \cite{nic1,nic2,nic3,riz,meh1})
\begin{equation}\label{mat:1}  \Bigg\{
\begin{array}{ll}
\rho_m(r)={M\over
{(4\pi\theta)^{\frac{3}{2}}}}e^{-\frac{r^2}{4\theta}}\\
 \rho_e(r)={Q\over {(4\pi \theta)^{\frac{3}{2}}}}
e^{-\frac{r^2}{4\theta}},\\
\end{array}
\end{equation}
where $\theta$ is the smallest fundamental cell of observable area
in the noncommutative coordinates, beyond which coordinate
resolution is vague. The solution of the Einstein equations
associated with the above smeared sources leads to the metric of
noncommutative RN black hole as follows {\footnote{Throughout the
paper, natural units are used so that $\hbar = c = G = k_B = 1$.}}
\begin{equation}
\label{mat:2}ds^2=-\bigg(1-\frac{2M_\theta}{r}+\frac{Q_\theta^2}{r^2}\bigg)dt^2+
\bigg(1-\frac{2M_\theta}{r}+\frac{Q_\theta^2}{r^2}\bigg)^{-1}dr^2+r^2
d\Omega^2,
\end{equation}
where $d\Omega^2=d\vartheta^2+\sin^2\vartheta d\varphi^2$ is the
line element on the 2-dimensional unit sphere. The smeared mass and
charge distributions are respectively given by
\begin{equation}\label{mat:3}  \Bigg\{
\begin{array}{ll}
M_{\theta}=M\left[{\cal{E}}\big(\frac{r}{2\sqrt{\theta}}\big)
-\frac{ r}{\sqrt{\pi\theta}}e^{-\frac{r^2}{4\theta}}\right]\\
 Q_\theta=Q\sqrt{{\cal{E}}^2\big(\frac{r}{2\sqrt{\theta}}\big)-
\frac{r}{\sqrt{2\pi\theta}}{\cal{E}}\big(\frac{r}{\sqrt{2\theta}}\big)}.\\
\end{array}
\end{equation}
In the limit of $\frac{r}{\sqrt{\theta}}\rightarrow\infty$, the
Gaussian error function defined as $ {\cal{E}}(x)\equiv
\frac{2}{\sqrt{\pi}}\int_{0}^{x}e^{-t^2}dt$, tends to one and we
recover the ordinary mass and charge perfectly localized at a point,
i.e. $M_\theta\rightarrow M$ and $Q_\theta\rightarrow Q$. Now, using
the Killing equation
\begin{equation}
\label{mat:4}\partial_a\xi_b+\partial_b\xi_a-2\Gamma^c_{ab}\xi_c=0,
\end{equation}
where $a,\,b,\,c\in\{0,\,1,\,2,\,3\}$. And the condition of static
spherical symmetry $\partial_0\xi_a=\partial_3\xi_a=0$, and also the
infinity condition $\xi_a\xi^a=-1$, the timelike Killing vector of
the noncommutative charged black hole can be written as
\begin{equation}
\label{mat:5}\xi_a=\left(\frac{2M_\theta}{r}-\frac{Q_\theta^2}{r^2}-1,\,0,\,0,\,0\right),
\end{equation}
which equals to zero at the event horizon. To illustrate a foliation
of space, and perceiving the holographic screens $\Omega$ at
surfaces of constant redshift, we consider the potential $\phi$ as
follows
\begin{equation}
\label{mat:6}\phi=\frac{1}{2}\log\left(-\xi^a\xi_a\right),
\end{equation}
we notice that $e^\phi$ implies the redshift factor and exhibits a
link between the local time coordinate and the reference point with
$\phi = 0$ at infinity.

\begin{figure}[htp]
\begin{center}
\includegraphics{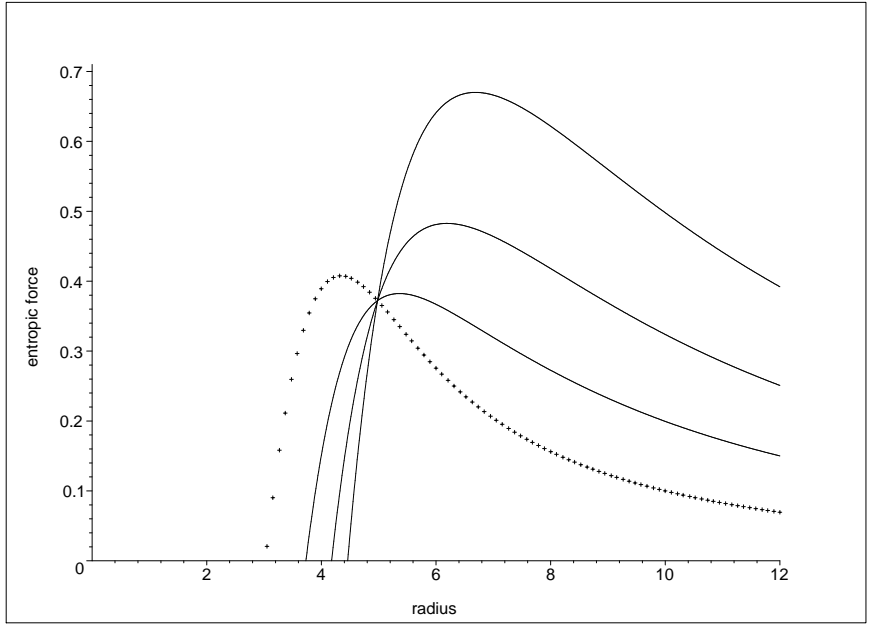}
\end{center}
\vspace{9.3 cm} \caption{\scriptsize {The entropic force $F$ versus
the radius, $\frac{r}{\sqrt{\theta}}$, for some values of
$\frac{Q}{\sqrt{\theta}}$. We have set $m=1.0\sqrt{\theta}$, and
$M=10.0\sqrt{\theta}$. On the right-hand side of the figure, from
bottom to top, the solid curves correspond to $Q =10.0\sqrt{\theta},
\, 15.0\sqrt{\theta}, \,$ and $20.0\sqrt{\theta}$, respectively. The
cross-dotted curve alludes to the noncommutative Schwarzschild black
hole so that it corresponds to $Q = 0$. According to the figure, the
appearance of a minimal nonzero radius, $r_0$, is evident.}}
\label{fig:1}
\end{figure}

\begin{figure}[htp]
\begin{center}
\includegraphics{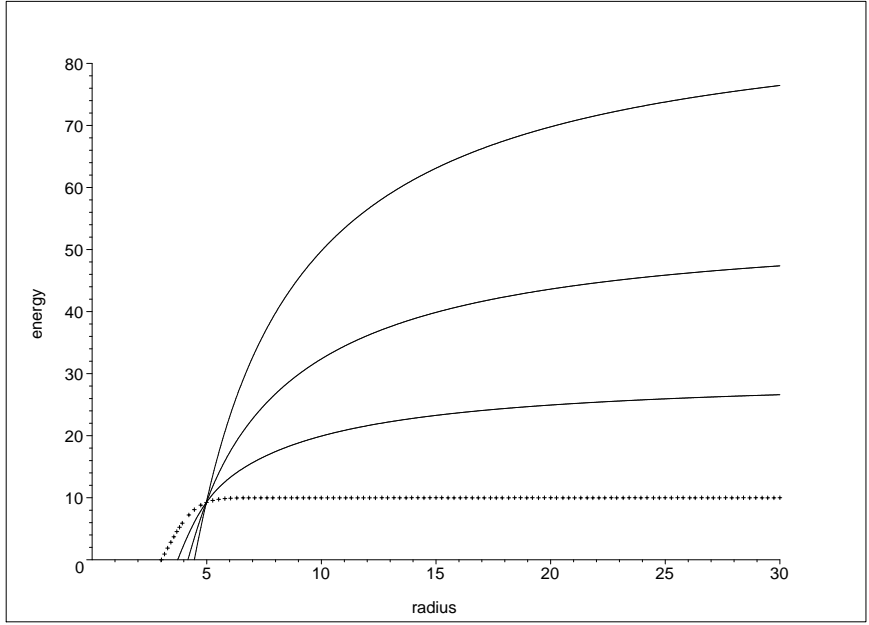}
\end{center}
\vspace{9.3 cm} \caption{\scriptsize {The energy,
$\frac{E}{\sqrt{\theta}}$, versus the radius,
$\frac{r}{\sqrt{\theta}}$, for some values of
$\frac{Q}{\sqrt{\theta}}$. We have set $m=1.0\sqrt{\theta}$, and
$M=10.0\sqrt{\theta}$. On the right-hand side of the figure, from
bottom to top, the solid curves correspond to $Q =10.0\sqrt{\theta},
\, 15.0\sqrt{\theta}, \,$ and $20.0\sqrt{\theta}$, respectively. The
cross-dotted curve alludes to the noncommutative Schwarzschild black
hole so that it corresponds to $Q = 0$.}} \label{fig:2}
\end{figure}

The acceleration $a^a$ on the spherical holographic screen with
radius $r$, for a particle which is situated extremely neighboring
the screen, is found to be \cite{ver}
\begin{equation}
\label{mat:7}a^a=-\nabla^a\phi.
\end{equation}
It is clear that the acceleration is perpendicular to the
holographic screen. By representing the normal vector as
$N^a=\frac{\nabla^a\phi}{\sqrt{\nabla^b\phi\nabla_b\phi}}$, the
local temperature on the screen is given by
\begin{equation}
\label{mat:8}T=-\frac{1}{2\pi}e^\phi N^a a_a.
\end{equation}
The above formula denotes that the four acceleration on the screen
is as follows: $a^a=\left(0,\,2\pi T,\,0,\,0\right)$, with a local
temperature carried by the noncommutative RN screen which is
computed in the following form:
\begin{equation}
\label{mat:9}T=\frac{M_{\theta}}{2\pi
r^2}-\frac{Mr}{4(\pi\theta)^{\frac{3}{2}}}e^{-\frac{r^2}{4\theta}}-\frac{1}{2\pi
r^3}\left[Q_\theta^2+Q^2\left(\frac{r^2}{2\pi\theta}-\frac{r}{\sqrt{\pi\theta}}
{\cal{E}}\Big(\frac{r}{2\sqrt{\theta}}\Big)\right)e^{-\frac{r^2}{4\theta}}\right].
\end{equation}
Note that the local temperature on the event horizon is identical to
the Hawking temperature, i.e., $T|_{r=r_{H}}=T_H$
\cite{nic1,nic2,nic3} (see also \cite{riz,meh1}). In the limit of
$\theta\rightarrow0$, one retrieves the standard temperature for the
RN case, i.e.
\begin{equation}
\label{mat:10}T_{RN}=\frac{M}{2\pi r^2}-\frac{Q^2}{2\pi r^3}.
\end{equation}
The alteration in entropy for a test particle with mass $m$ at fixed
place nearby the screen is equal to
\begin{equation}
\label{mat:11}\nabla_a S=-2\pi m N_a.
\end{equation}
Eventually, the modified Newtonian force law as the entropic force
in the presence of the noncommutative RN black hole can now be
written as
\begin{equation}
\label{mat:12}F=\sqrt{g^{ab}F_a
F_b}=\frac{mM_{\theta}}{r^2}-\frac{mMr}{2\sqrt{\pi\theta^3}}e^{-\frac{r^2}{4\theta}}
-\frac{m}{r^3}\left[Q_\theta^2+Q^2\left(\frac{r^2}{2\pi\theta}-\frac{r}{\sqrt{\pi\theta}}
{\cal{E}}\Big(\frac{r}{2\sqrt{\theta}}\Big)\right)e^{-\frac{r^2}{4\theta}}\right],
\end{equation}
where $F_a=T\nabla_a
S=\left(0,\,\frac{m}{2\sqrt{g_{00}}}{{dg_{00}}\over
{dr}},\,0,\,0\right)$. If we choose the noncommutative Schwarzschild
case, i.e. $Q=0$, then we have
\begin{equation}
\label{mat:13}F=\frac{mM_{\theta}}{r^2}-\frac{mMr}{2\sqrt{\pi\theta^3}}e^{-\frac{r^2}{4\theta}}.
\end{equation}
For the ordinary RN case, the entropic force becomes \cite{liu,cha}
\begin{equation}
\label{mat:14}F_{RN}=2\pi mT_{RN}=\frac{mM}{r^2}-\frac{mQ^2}{r^3}.
\end{equation}
The numerical results of the entropic force versus the radius, for
several values of $\frac{Q}{\sqrt{\theta}}$, are displayed in
Fig.~\ref{fig:1}. This figure shows that the peak entropic force
drops with decreasing the electric charge. In accord with the
figure, in the frame of noncommutative geometry inspired model, the
entropic force of the black hole becomes larger with the reduction
of the screen radius up to the time when it comes near to a highest
definite value and afterwards goes down to zero at the minimal
nonzero value of the screen radius, $r_0$. In fact, on account of
coordinate noncommutativity, the black hole entropic force falls
down to zero at the remnant size (does not diverge at all), and
therefore ceasing to exist the
divergence is obvious.\\
Assuming that the mass of the source becomes larger than the mass of
the test particle and is placed at the origin of coordinate, we
suppose the energy connected to the source, including
noncommutativity effects, is dispersed on a closed screen of
constant redshift $\phi$. On this surface, $N$ bits of data are
accumulated and the holographic data from the source is encrypted as
$dN=dA$ \cite{pad2}, where $A$ is the area of the surface. The
energy on the noncommutative RN screen, in agreement with the
Gauss's theorem, approves thermal equipartition,
\begin{equation}
\label{mat:15}E=\frac{1}{2}\int_\Omega TdN=\frac{1}{4\pi}\int_\Omega
e^\phi\nabla\phi dA.
\end{equation}
Thus, we have
\begin{equation}
\label{mat:16}E=M_{\theta}-\frac{Mr^3}{2\sqrt{\pi\theta^3}}e^{-\frac{r^2}{4\theta}}
-\frac{Q_\theta^2}{r}-Q^2\left(\frac{r}{2\pi\theta}-\frac{{\cal{E}}\big(\frac{r}{2\sqrt{\theta}}\big)}{\sqrt{\pi\theta}}
\right)e^{-\frac{r^2}{4\theta}}.
\end{equation}
If we consider the case of $Q=0$, then we can obtain the following
relation for the energy on the noncommutative Schwarzschild screen:
\begin{equation}
\label{mat:17}E=M_{\theta}-\frac{Mr^3}{2\sqrt{\pi\theta^3}}e^{-\frac{r^2}{4\theta}}.
\end{equation}
For the commutative case, $\theta\rightarrow0$, the energy on the
ordinary RN screen is as follows \cite{liu}:
\begin{equation}
\label{mat:18}E_{RN}=2\pi r^2T_{RN}=M-\frac{Q^2}{r}.
\end{equation}
The plot presented in Fig.~\ref{fig:2} shows the numerical results
of the energy versus the radius, for several values of
$\frac{Q}{\sqrt{\theta}}$. Fig.~\ref{fig:2} clearly shows that for a
very large value of the screen radius,
$\frac{r}{\sqrt{\theta}}\gg1$, the energy on the screen will be
constant and the constant peak of the energy increases with
enlarging $\frac{Q}{\sqrt{\theta}}$. The disappearance of divergence
for the energy on the screen, because of the presence of the
residual nonzero size of the black hole, can also be demonstrated by
decreasing the value of the screen radius. According to the figure,
as electric charge becomes larger the minimal nonzero radius
increases; this corresponds to Fig.~\ref{fig:1}.

As can be seen from these two figures, the entropic force and the
energy on a holographic screen with radius $r_0$ are zero. This is a
prominent consequence which means that since $r_0$ is the radius of
smallest holographic screen, it cannot probe via a test particle
which is placed within a short distance from the source. Therefore
the standard formulation of Newtonian gravity, in very short
distances when the screen radius reaches to $r_0$, is broken down.
This means that the test particle with mass $m$ cannot identify any
gravitational field, when it is at a minimal distance $r_0$ from the
source mass. This phenomenon violates the existence of the mere
gravitational interaction for an inert residue of the black hole.
The black hole remnant as an indispensable physical object is
greatly approved in the quantum gravity literature when quantum
gravitational fluctuations are revealed. For instance, when
generalized uncertainty principle is taken into account, the
complete decay of the black hole into emission is not allowed; then
there would be a massive and inert remnant containing the sole
gravitational interactions \cite{adl}. Our method manifestly
demonstrates that the residue of the black hole is entirely inert
without any gravitational interaction included. From another point
of view, the EP which is relevant to the equality of gravitational
and inertial mass, is contravened because it is now conceivable to
distinguish between them. Indeed, the gravitational mass in the
leftover size of the inert black hole does not radiate any
gravitational field, hence it is identified to be zero as opposed to
the inertial mass.\\

In summary, we have discussed some aspects of Verlinde's proposal in
the presence of noncommutative charged black hole based on
Gaussian-smeared mass distribution. We have considered the case of
noncommutative geometry inspired Reissner-Nordstr\"{o}m black hole
to improve the expression of black hole's entropic force by taking
into consideration the noncommutative corrections. In this setup, we
have exhibited that there is a violation of the EP when we apply the
noncommutativity effects in our computations, i.e. one can
distinguish between a uniform acceleration and a gravitational
field. In conclusion, it seems that the noncommutative gravity
effects exhibit a conflict with the entropic idea of Newtonian
gravity. However, it is possible to assume that both entropic
gravity and noncommutative geometry can be correct but in short
distances one predicts a violation of EP. This means that it may be
feasible that when one approaches $r_0$ one diverges from GR. Hence,
there can be a prediction that EP is contravened at some small scale
(maybe the Planck scale) caused by the combination of entropic
gravity and noncommutative geometry.

\end{document}